\documentclass{iau} 

\usepackage{graphicx}
\usepackage{xcolor}
\usepackage[hyphens]{url}
\usepackage{hyperref}
\hypersetup{
     colorlinks  = true,
     linkcolor   = blue, 
     anchorcolor = BrickRed,
     citecolor   = blue,
     filecolor   = blue,
     menucolor   = blue,
     runcolor    = cyan,
     urlcolor    = blue
}
\usepackage[authoryear]{natbib}

\pubyear{2019}
\volume{355} 
\jname{The Realm of the Low-Surface-Brightness Universe}
\editors{D. Valls-Gabaud, I. Trujillo \& S. Okamoto, eds.}
\def\etal{\textit{et al.}}
\def\magsec{mag arcsec$^{-2}$}
\def\ha{H$\alpha$}
\def\haunits{${\rm erg\ s^{-1}\ cm^{-2}\ arcsec^{-2}}$}

\title[Deep Imaging of Galaxies and Clusters] %% give here short title %%
{Deep Imaging of Diffuse Light\\ Around Galaxies and Clusters:\\ Progress and Challenges}

\author[Chris Mihos]   %% give here short author list %%
{J. Christopher Mihos$^1$}

\affiliation{$^1$Department of Astronomy\\ Case Western Reserve University \\ 10900 Euclid Ave,
Cleveland OH, USA\\ email: {\tt mihos@case.edu}}

\begin{document}

\maketitle

\begin{abstract}
Over the past several decades, advances in telescope/detector technologies and deep imaging
techniques have pushed surface brightness limits to ever fainter levels. We can now both detect
and measure the diffuse, extended star light that surrounds galaxies and permeates galaxy
clusters, enabling the study of galaxy halos, tidal streams, diffuse galaxy populations, and the
assembly history of galaxies and galaxy clusters. With successes come new challenges, however,
and pushing even deeper will require careful attention to systematic sources of error. In this review
I highlight recent advances in the study of diffuse starlight in galaxies, and discuss challenges
faced by the next generation of deep imaging campaigns.
%%%%%%%%%%%%%%%%%%%%%%%%%%%%%%%%%%%%%%%%%%%%%%%
\keywords{galaxies: general, galaxies: clusters: general, telescopes, surveys}
\end{abstract}

\firstsection % do not remove

\section{Introduction}

While much attention has been focused on extreme low surface brightness
(LSB) science\footnote{Here I rather arbitrarily define ``extreme LSB''
as optical surface brightnesses at or below 1\% of the ground-based
night sky brightness, or $\mu_V \gtrsim 26-27$ \magsec.} in recent
years, the field itself is hardly new. As far back as eighty five years
ago, \citet{Stebbins34} were using a photoelectric photometer on the Mt
Wilson 100'' telescope to trace M31's surface brightness profile down to
$\mu_{pg} \approx 26-27$ \magsec, while by the 1950s Zwicky was already
postulating the existence of intracluster light (ICL) based on imaging
of tidal streams around cluster galaxies \citep{Zwicky52}. Indeed, many
of the scientific issues this conference focuses on have been motivated
by deep imaging studies in the 1970s and 1980s. Deep photographic
imaging revealed the structure and color of the extended envelopes of
bright ellipticals \citep[e.g.,][]{Arp69,deVauc69}, and showed a myriad
of stellar shells and streams marking the accretion of material onto
spiral and elliptical galaxies \citep[e.g.,][]{Malin80,Schweizer88}. On
larger scales, the diffuse ICL in Coma was first imaged in the early
1970s \citep{deVauc70, Welch71}, while deep surveys of Virgo in the
1980s had already uncovered a population of large and extremely diffuse
galaxies throughout the cluster \citep{Sandage84}.

In the years since these discoveries, advances in telescope technology,
instrumentation, and data analysis have allowed astronomers to survey
the low surface brightness universe down to even greater depths and over
wider areas than previously possible. Modern imaging that carefully
corrects for (or mitigates) contamination due to stray light and
astronomical foregrounds and backgrounds can reach down to optical
surface brightnesses as low as 29--30 \magsec\ --- and even lower in
certain cases. We have now moved beyond simple detection of LSB features
and into the realm of accurate photometric studies of the structure and
stellar populations that compose the diffuse starlight surrounding
galaxies.

Deep LSB imaging of galaxies and clusters is particularly important
given the information held in the low surface brightness structure that
surrounds them. The dynamical timescales are long in galaxy outskirts,
such that the faint tidal streams and shells from past interactions and
accretion events may survive for many Gyr. The morphologies, colors, and
kinematics of tidal tails provide important constraints on the dynamical
evolution of interacting galaxies. Deep imaging can also reveal the
structure of diffuse galaxy halos and the history of satellite accretion
in galaxies. On the largest scales, the diffuse intracluster light (ICL)
that permeates massive galaxy clusters can be used to trace cluster
assembly history. They key requirement, though, is to go deep: it is at
the lowest surface brightnesses ($\gtrsim$ 28--30 \magsec) that galaxy
halos are predicted to be awash in accretion streams, and where the
diffuse light in clusters decouples from the galaxy outskirts and
instead traces the complex accretion structure of the intracluster
light.

\section{Deep Imaging: Depths and  Strategies}

Before embarking on an overview of LSB imaging studies, it is important
to recognize that, unlike point source depths, there is no well-defined
metric for surface brightness depth; this has led to different groups
quantifying depths in different ways. Complicating matters is the fact
that some studies report limiting depths in terms of derived quantities:
surface brightnesses of detected objects, limits to surface brightness
or color profiles, etc. Comparisons such as these conflate the quality
of the {\sl data} and the depth of the {\sl analysis}. Even the most
straight-forward metric --- the background variation in the imaging
measured over a fixed spatial scale \citep[\eg][]{Trujillo16,Mihos17}
--- is not uniquely defined, as it employs a scale that will necessarily
differ depending on the characteristics of the telescope and detector.
For example, {\sl Hubble ACS} imaging has 0.05$^{\prime\prime}$ pixels
and a $202^{\prime\prime}{\times}202^{\prime\prime}$ FOV, while small
ground-based telescopes have arcsecond-scale pixels and degree-scale
FOVs. When both the pixel scale and areal coverage differ by more than
an order of magnitude between two datasets, a direct statistical
comparison of the background fluctuations on fixed angular scale is
ill-posed, and ultimately not well-motivated. The scientific goals of
such projects are typically quite different, as {\sl Hubble} studies of
diffuse light have focused on higher redshift systems with small angular
scale, while LSB imaging of nearby galaxies has been the province of
ground-based observatories with wide-field capabilities.

In addition, if the goal is to go uniformly deep over a wide area,
metrics should also quantify both the limiting depth and its variation
over large angular scales, a practice that is rarely done. Here too,
however, singular definitions can be problematic. Studies that search
for small-scale diffuse features (LSB galaxies or tidal streams around
distant objects) can tolerate large-scale variations in depth as long as
a local background can be accurately modeled. In contrast, when studying
the extended halos of nearby galaxy or diffuse light in clusters, the
imaging {\sl must} be very uniform over much larger scales. All these
issues make direct comparisons of published depths quite
difficult.\footnote{The task is made even {\sl more} difficult by
sometimes hyper-competitive focus on being ``The Deepest.'' This has
lead to claims of limiting depths which are often over-stated, or the
use of improper comparison metrics cherry-picked to advantage a
particular study. This practice distorts the science and is harmful to
the field.} As a result, in this review I make no attempt to
``standardize'' depths to a single unique metric, but instead give the limiting
depths as reported by different surveys, highlighting the angular scale
and statistical metric used whenever possible. Ultimately I leave
the reader with one final warning when interpreting limiting depths:
{\sl caveat emptor}.

Over the past years, a variety of deep imaging projects have come online
which employ different strategies to probe the low surface brightness
universe. One approach uses small telescopes or telescope arrays
optimized for low surface brightness imaging. Such systems are
competitive even in the world of large telescopes, as photometric errors
in deep surface photometry are dominated not by photon statistics but by
systematic sources of uncertainty due to flat-fielding variations, sky
and background estimation, and contamination by stray light. This latter
issue is perhaps the most problematic, and arises from a variety of
sources including the extended wings of the PSF, extraneous off-axis
light, scattered light within the telescope, reflections between various
optical elements, and complex diffraction patterns from obstructions in
the telescope beam. Thus, systems which mediate these effects through
optical design choices (using closed tube telescopes or telescopes with
unobstructed beams, reducing the number of reflective surfaces,
employing aggressive anti-reflective coatings, etc) can provide
significant advantages. Furthermore, for surface photometry, the fast
beams and large pixel scales of these systems provide additional
benefits: more photons per pixel to estimate sky backgrounds and reduce
photon noise, and the ability to image large fields of view without the
need for CCD arrays and the chip-to-chip sensitivity variations they
introduce. Examples of these types of telescopes include the
LSB-optimized Burrell Schmidt telescope \citep{Mihos17} reaching
limiting depths of $\mu_{B,lim} \approx 29.5$ \magsec\ ($3 \sigma, 1'$
scales), and the Dragonfly imaging array \citep{Abraham14} which reaches
a comparable depth of $\mu_{g,lim} \approx 29.5$ \magsec\ \citep[$1
\sigma, 1'$ scales;][]{Merritt16}.

\begin{figure}[!h]
\begin{center}
 \includegraphics[width=1.0\textwidth]{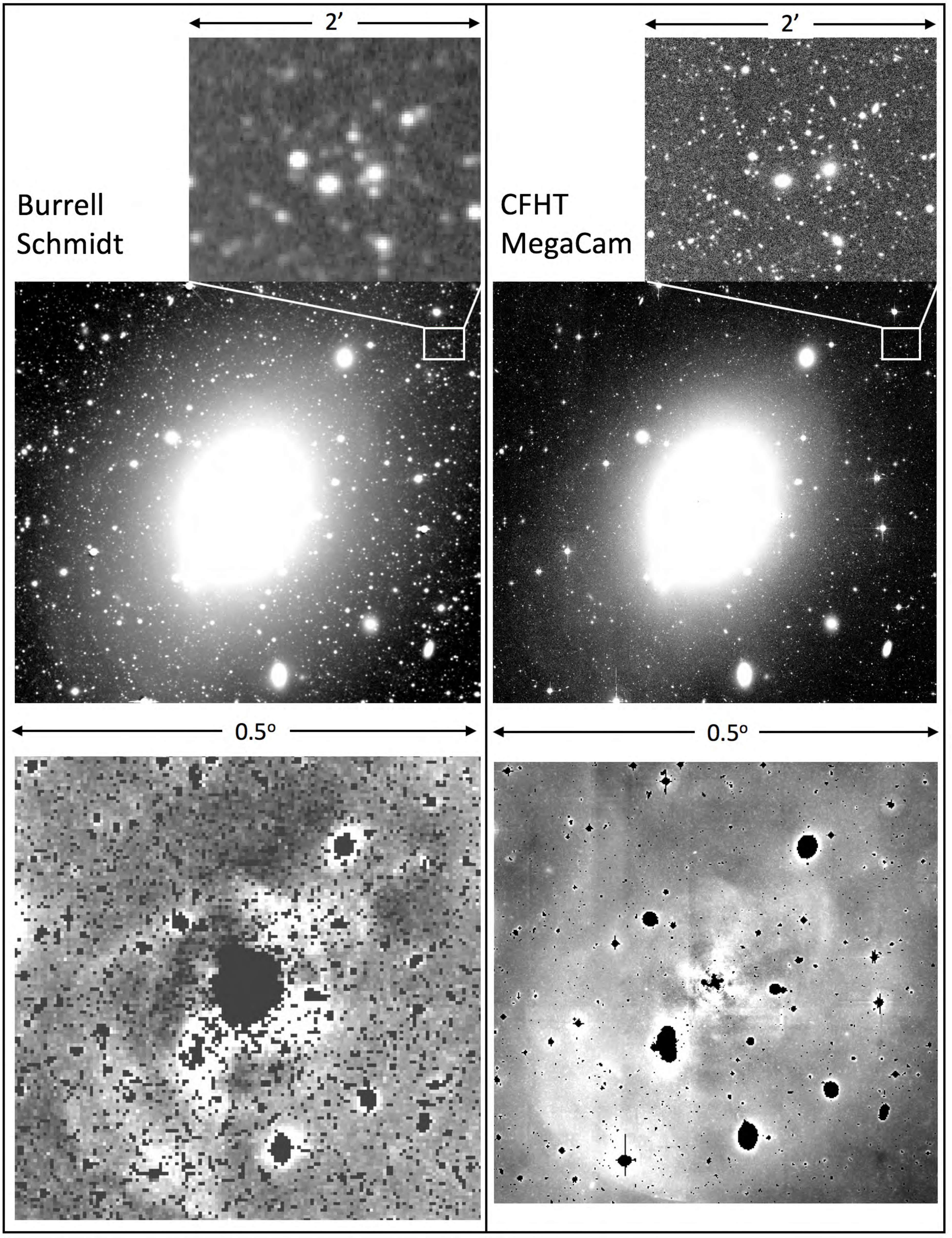} 
\caption{Deep imaging of the Virgo elliptical galaxy M49 from the
0.6/0.9m Burrell Schmidt (left, from \citealt{Janowiecki10}) and the 3.5m
CFHT (right, from \citealt{Arrigoni12}) telescopes. The center images span
0.5$^\circ$ across, while the small images at the top show a 2$'$ cutout
of each image. The bottom images show the result of subtracting a smooth
model for the M49 light, then masking bright pixels and spatially
re-binning the images to show residual low surface brightness features.}
  \label{largesmall}
\end{center}
\end{figure}

Of course larger telescopes deliver significant advantages of their own,
most obviously in their collecting area but also in finer pixel scale.
This provides not only more photons and better spatial resolution, but
the ability to better resolve and mask out stars and background galaxies
that would otherwise contaminate the surface photometry. Large
telescopes equipped with large format CCD arrays can now image over a
wide area, albeit often at the cost of increased scattered light from
correcting optics and/or complicated sensitivity variations across the
array. These tradeoffs are illustrated in Figure~\ref{largesmall}, which
compares deep imaging of the Virgo giant elliptical M49 using the
0.6/0.9m Burrell Schmidt telescope \citep{Janowiecki10} to that taken
with Megacam on the 3.6m CFHT telescope \citep{Arrigoni12}. The center
panels show the raw imaging data, with the top insets showing a 2$'$
zoom of a portion of the field. Compared to the Burrell Schmidt imaging,
the superior resolution and point source depth of the CFHT data is
clear. The bottom panels shows the result after subtracting smooth
models for M49's light, masking compact sources, and median-binning the
images spatially to highlight M49's diffuse shell system. In the CFHT
imaging, the higher spatial resolution and better masking of
contaminants leads to a smoother, more detailed map of the shells on
small scales. However, on larger scales chip-to-chip variations in
sensitivity can be seen as linear artifacts in the residual image, along
with circular reflections from bright stars in the field. In contrast,
the Burrell Schmidt imaging is more uniform on large scales, due to its
single CCD detector and use of aggressive anti-reflection coatings on
the optical elements of the camera.

Deep targeted surveys using 3m-class telescopes have studied nearby
galaxies and clusters, including the {\sl Next Generation Virgo Cluster
Survey} \citep{Ferrarese12} and ATLAS3D survey \citep{Cappellari11} on
CFHT, and the {\sl Fornax Deep Survey} \citep{Iodice16} and VEGAS survey
\citep{Capaccioli15} on the VST telescope at ESO. While these telescopes
are not specifically optimized for LSB science, through careful
reduction techniques these surveys have typically achieved limiting
surface brightnesses of $\mu_{g,lim} \approx 28.5-29$ \magsec.
Complementary to these targeted surveys are new wide-area imaging
surveys such as the {\sl CFHT Legacy Survey} \citep{Gwyn12}, the {\sl
DECam Legacy Survey} on the CTIO 4m \citep{Dey19}, and the {\sl Hyper
Suprime-Cam Subaru Strategic Program} on the 8.2m Subaru telescope
\citep{Aihara18}. Compared to targeted surveys, these wide-area surveys
are somewhat shallower ($\mu_{r,lim} \approx$ 27.5--28.5 \magsec;
\citealt{Atkinson13,Greco18,Hood18}) but they cover hundreds to
thousands of square degrees of the sky, enabling the study of LSB
structures around large samples of galaxies spanning a wide range of
physical environments.

The potential to push extraordinarily deep with large telescopes was
demonstrated by \citet{Trujillo16}, who used the 10m GTC telescope to
push down to a limit of $\mu_{r,lim} \approx 31.5$ \magsec, albeit over
only a small field of view ($< 5'$). Over larger areas such depths are
not yet attainable, but looming on the horizon is LSST, whose 8.4m
primary and 9.6 deg$^2$ camera will repeatedly image the night sky
beginning in 2022. When completed, the main survey will deliver 825
dithered exposures of each patch of sky within its 18,000 deg$^2$ survey
area. These dithered images will be critical for LSB science, not just
for stacking to build signal, but also for identifying and removing
scattered light in the imaging. While forecasting surface brightness
depth is a dangerous game, if scattered light can be properly mitigated
or modeled out of the data (see \S4.1 below), LSST could potentially
reach depths of $\mu_{g,lim} \approx 28$ \magsec\ in a single visit, and
$\mu_{g,lim} \approx 31$ \magsec\ in a full stack of dithered images
\citep{Laine18}.

Space-based observatories offer advantages to LSB science as well,
including more compact PSFs, lower sky backgrounds, and ultraviolet
capabilities. Several deep optical imaging campaigns using {\sl Hubble}
have delivered rich datasets with the potential to reach very low
surface brightness limits, such as the {\sl Hubble Ultradeep Field} and
its descendants \citep{Beckwith06,Koekemoer13,Illingworth13} and the
{\sl Hubble Frontier Fields} \citep{Lotz17}. Recent work studying the
diffuse light in Frontiers Field clusters \citep{Montes14, Morishita17,
Montes18} and extended light of galaxies in the Ultradeep Field
\citep{Borlaff19} have demonstrated that these datasets can reach limits
as deep as $\mu \approx 31$ \magsec\ ($3 \sigma, 3''$ scales). However,
to fit in {\sl Hubble}'s small field of view, target galaxies and
clusters must be at higher redshift, where cosmological surface
brightness dimming makes the already-faint diffuse light around them
even fainter. Fortunately, upcoming space missions such as {\sl WFIRST}
and {\sl Euclid} will have much wider fields of view, potentially
delivering LSB capabilities even for nearby galaxies and clusters.
Greater gains may be possible through the development of space
telescopes optimized for LSB imaging. One such example is the proposed
MESSIER surveyor \citep{Valls17}: a small (50cm) and fast ($f/2$) space
telescope using a bi-folded Schmidt optical design to eliminate the
extended PSF wings that normally arise from obscuration by the secondary
\citep{Muslimov17}. With a $2^\circ \times 4^\circ$ FOV and optimized to
work in both optical and ultraviolet, MESSIER would conduct an all-sky
survey probing diffuse starlight around galaxies, identifying the most
diffuse galaxy populations, and tracing Ly$\alpha$ emission from the
cosmic web.

\section{Recent LSB Science Highlights}

Here I highlight a handful of studies which demonstrate the
state-of-the-art capabilities in LSB studies of diffuse light around
galaxies and clusters. I concentrate on results gleaned from imaging in
{\sl integrated light}; additional {\sl critical} information comes from
spectroscopic data, studies of resolved stellar populations, and the
properties of discrete tracers such as PNe and globular clusters, but
will not be discussed here.

\subsection{Stellar Populations in Galaxy Outskirts}

The colors of the diffuse light surrounding galaxies provide important
constraints on the stellar populations in their outer disks and halos.
While broadband colors are a relatively crude tool for stellar
population work, suffering from the well-known age-metallicity
degeneracy, in the majority of cases they are the only means available
for studying populations at low surface brightness. Compared to
broadband imaging, spectroscopic studies become prohibitively difficult
at extremely low surface brightness, while resolved star observations,
perhaps the ``gold standard'' for studying stellar populations, are
currently impractical beyond 10--15 Mpc. Because so much of our
understanding of stellar populations is based on resolved star studies
in the Local Volume, while studies of more distant galaxies must rely on
imaging in integrated light, it is particularly important to cross-check
inferences from the two techniques in galaxies where both are possible.

The halos of nearby spiral galaxies provide one such opportunity.
Because their integrated light is dominated by old red giant branch
stars, halo stellar populations are relatively simple and make for a
clean test of the two techniques. But photometric accuracy in integrated
light is critical; color uncertainties even as small as 0.1 mag can lead
to significant uncertainty in the underlying populations. Indeed, some
early work tracing the red colors of outer disks and halos were unduly
influenced by the extended wings of the PSF scattering light outwards
from the bright nuclei of the galaxies \citep[see \eg\ the discussion
in][]{Sandin14,Sandin15}. However, recent studies do much better at
constraining colors at low surface brightness. For example, deep imaging
of the halo of NGC 4565 by Infante-Sainz et al. (in prep) achieves color
uncertainties of $\lesssim 0.05$ mag at $\mu_r\approx28$ \magsec, and
indicates metallicities in the range [Fe/H] = $-0.7$ to $-1.7$. This
compares favorably to the resolved star analysis of \citet{Monachesi16},
which derives a halo metallicity of [Fe/H] = $-1.2\pm0.3$ for fields at
similar radii. The consistency in the inferred metallicity is promising,
although the quality of the match is still sensitive to systematic
details (such as the population age) in the modeling of the integrated
colors.

\begin{figure}[!h]
\begin{center}
 \includegraphics[width=1.0\textwidth]{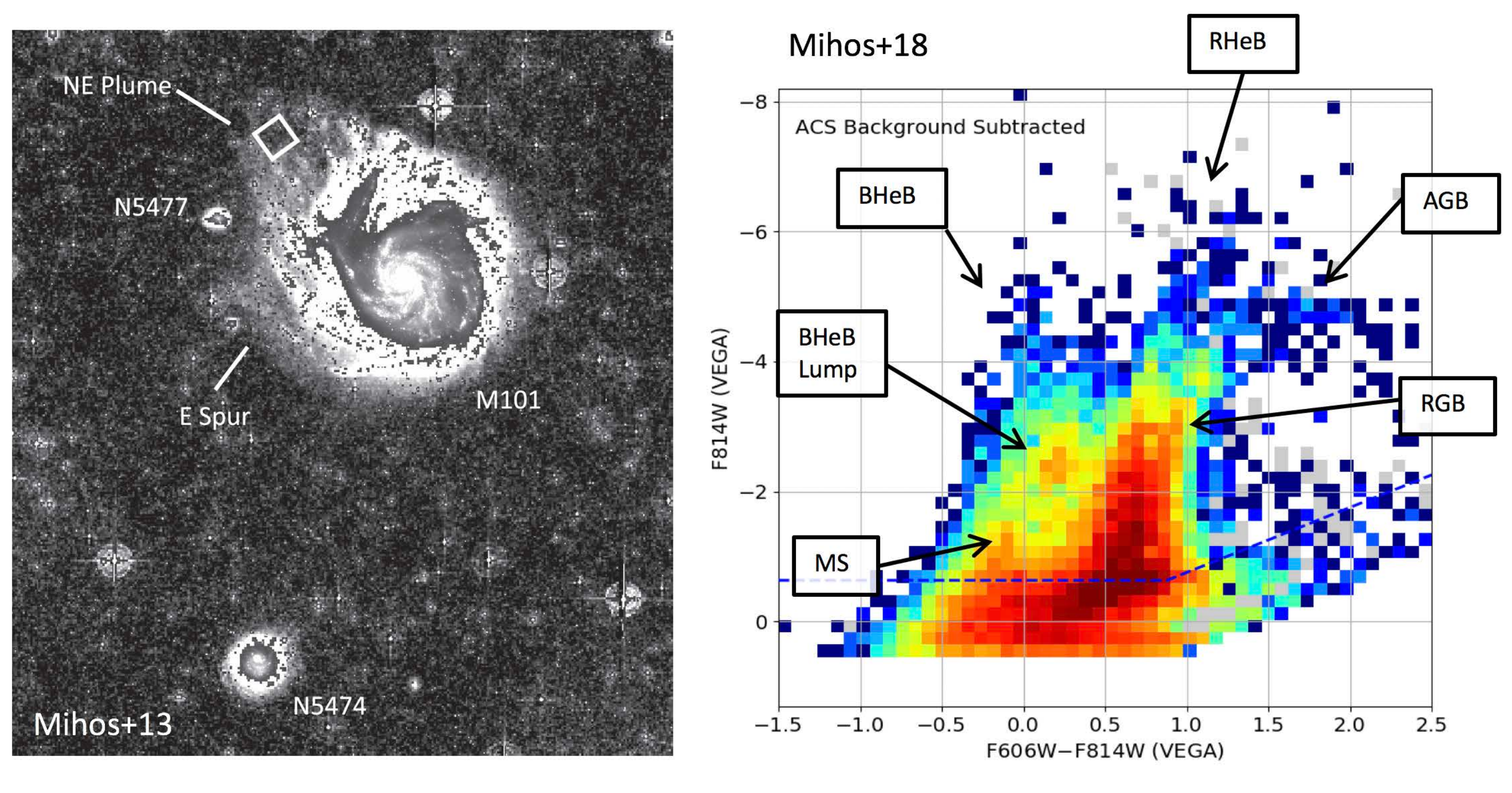} 
\caption{Left: Deep imaging of the spiral galaxy M101 showing the
diffuse outer regions of the disk \citep{Mihos13}. The white
box in the NE Plume shows the location of our {\sl Hubble} ACS imaging.
Right: Color-magnitude diagram of stars in {\sl Hubble} imaging, showing 
various stellar populations including a coeval lump of stars in
the BHeB sequence.
 \citep{Mihos18}.}
  \label{M101}
\end{center}
\end{figure}

A second test involves the star formation history in outer disks; here,
the situation becomes more complex, as multiple stellar populations
contribute to the integrated light. Nonetheless, significant headway can
be made by folding in additional constraints. Deep imaging of the nearby
spiral M101 by our group \citep[][see Figure~\ref{M101}a]{Mihos13}
revealed the galaxy's distorted outer disk extending nearly 50 kpc from
the center. The very blue colors of the light ($B-V=0.2\pm0.05$ at
$\mu_B = 29.5$ \magsec), coupled with the lack of significant {\it
GALEX} FUV emission, argued for a weak burst of star formation 250--350
Myr ago which has since largely died out. Subsequent {\sl Hubble}
imaging \citep{Mihos18} resolved multiple stellar populations
(Figure~\ref{M101}b), including a discrete ``lump'' of stars in the blue
helium burning sequence indicative of a coeval population evolving off
the main sequence. Detailed stellar population models place the age of
this population at 300--400 Myr, in excellent agreement with constraints
from the integrated light. Of course, the resolved populations provide
additional constraints as well, showing the stellar populations of the
outer disk to be quite metal-poor, with [Fe/H] = $-1.15\pm0.2$.

The agreement between the resolved populations and integrated light in
studies such as these give confidence in our ability to study stellar
populations at low surface brightness. However, the number of galaxies
in which these tests have been done is relatively small. As upcoming
missions like {\sl JWST, WFIRST, {\rm and} Euclid} extend both the reach
and scope of resolved population studies, and as more deep imaging of
diffuse integrated light becomes available from existing telescopes and
LSST, comparing these techniques in populations which span a wider range
of environment and galaxy type will be particularly important.

\subsection{Galaxy Accretion and the Structure of Stellar Halos}

At sufficiently low surface brightness, the halos of galaxies should be
filled with streams from accreted satellites. The morphologies of these
streams are sensitive to properties of the accretion, such as mass and
angular momentum of the satellite, and the time of accretion
\citep{Johnston16,Karademir19}. These events are well-traced by resolved
star imaging of galaxies within 5 Mpc \citep[\eg][]{Okamoto15,
Crnojevic16, McConnachie18}, and many more tidal streams have been found
in the integrated light around more distant galaxies. \citep[see,
\eg,][]{MartinezDelgado19}. As new wide-area surveys probe down to $\mu
\approx 28$ \magsec\ and below, tidal streams can be detected around
much larger samples of galaxies, studying the accretion properties of
galaxies as a function of morphology, mass, color, and environment.

Traditionally, identification and classification of accretion events in
deep imaging has largely been done by eye \citep[\eg][]{Nair10,
Atkinson13, Hood18}, as the low surface brightness and irregular
structure of tidal streams makes them particularly challenging targets
for automated detection algorithms. The quantity of data due in from new
deep imaging surveys make this approach impractical; instead, new
algorithms are needed to detect accretion signatures around much larger
samples of galaxies. This is a challenging task. For example, while the
algorithm of \citet{Kado-Fong18} identified high-frequency substructure
around galaxies, morphological {\sl classification} of these structures
still required human inspection. In contrast, \citet{Walmsley19} have
developed a neural network algorithm to both identify and classify tidal
debris around galaxies in deep imaging but the technique is limited by
the lack of comprehensive training samples. Ideally, algorithms would
connect the morphology of the detected features back to the physical
parameters of the accretion event: angular momentum, mass, accretion
time, etc. Some progress on this front has been made by
\citet{Hendel19}, who test automated morphology algorithms against
controlled N-body simulations of accretion events. While all these
techniques show promise, training and applying them across large and
heterogeneous survey datasets remains a work in progress.

At even lower surface brightness, more information is available in
galaxies' diffuse stellar halos. First, many more tidal streams are
thought to be found at or below $\mu_g \sim $ 29 \magsec\
\citep{Bullock05}; the bright streams found to date represent only the
tip of the iceberg. Over time, these streams will mix spatially and
become even more diffuse, but even once streams are lost as individual
objects, simulations suggest that the structure of the smooth stellar
halo remains sensitive to galaxies' accretion properties, with steeper
density profiles and metallicity gradients reflecting a quieter
accretion history \citep{Cook16}. However, tracing these halos requires
imaging well below $\mu_g =$ 30 \magsec, an extremely challenging
prospect for current imaging campaigns. Recently, \citet{Merritt16} used
Dragonfly imaging with characteristic depth $\mu_g \approx 29.5$
\magsec\ ($1\sigma, 1^\prime$ scale) to search for halo light in a
sample of nearby spirals. While azimuthal averaging erases spatial
information in the imaging, it allows one to push to lower surface
brightness, and the \citeauthor{Merritt16} analysis traced the
azimuthally averaged surface brightness profiles down to $\mu_g \approx
30-32$ \magsec\ ($2\sigma$). At this level, the profiles suggest a wide
range of inferred halo properties --- including several galaxies with
{\sl no} detected halo light --- and provide some tension with
cosmological simulations of galaxy formation. But with only eight
galaxies in the sample it is difficult to draw broad conclusions; a
proper understanding of the structure and demographics of galaxy halos
will require deeper imaging across much larger samples.

\subsection{Intracluster Light}

The diffuse intracluster light (ICL) that pervades galaxy clusters
traces galaxy accretion over the largest scales. Mapping the ICL in
nearby clusters like Virgo and Fornax is particularly important, because
at this distance we have complementary information on ICL stellar
populations and kinematics from discrete tracers such as RGB stars,
planetary nebulae, and globular clusters. However, nearby clusters are
large in angular size (Virgo's core [virial] radius spans $\sim
2.1^\circ\ [5.4^\circ]$; \citealt{Ferrarese12}) such that LSB imaging of
these clusters demands deep {\sl and uniform} imaging over very wide
fields. Much of the core of Virgo (along with that of the M49 subgroup)
was first imaged down to $\mu_V=28.5$ \magsec\ by
\citet{Mihos05,Mihos17}, with the whole 105 deg$^2$ virial area of the
cluster subsequently imaged to comparable depths by the {\sl Next
Generation Virgo Survey} team \citep{Ferrarese12}. In the south, the
{\sl Fornax Deep Survey} has mapped the diffuse light in Fornax
\citep{Iodice16}, although reflections from bright foreground stars
complicate the imaging in many areas of the cluster.

These surveys, along with imaging of more distant clusters, have yielded
a plethora of information about diffuse galaxy populations and ICL
formation. The large low surface brightness galaxies first identified in
Virgo by \citet{Sandage84} have been uncovered in large numbers in Coma
\citep[and re-dubbed ``ultra-diffuse galaxies'';][]{vanDokkum15, Koda15}
and Fornax \citep{Venhola17}, and are now detected at even fainter
levels in Virgo \citep{Mihos05,Mihos17}. Merging clusters show
cluster-wide plumes of diffuse light \citep[\eg][]{Feldmeier04,
Arnaboldi12}, evidence of rapid, on-going generation of ICL during
cluster mergers. On smaller scales, faint streamers and shells in the
halos of cluster ellipticals trace the tidal stripping of galaxies in
the cluster environment \citep{Janowiecki10}. The halos of massive Virgo
and Fornax ellipticals have been imaged out beyond 100 kpc
\citep[\eg][]{Rudick10, Mihos13, Iodice16}; at these radii, the light
become significantly bluer, indicative of younger and/or more metal-poor
stellar populations where the halo begins to merge into the cluster's
diffuse ICL. The long, thin ICL streams in Virgo also have colors
similar to those of low luminosity spheroidal galaxies in the cluster
\citep{Rudick10}, arguing that tidal shredding of faint galaxies is
another important channel for ICL formation.

As clusters assemble, the fraction of light in the ICL is expected to
grow, and its morphology will evolve as streams continually form and
disrupt \citep{Rudick06}. Imaging clusters at higher redshifts thus
affords the possibility of tracking the evolution of ICL with time.
Several recent studies have used the deep Hubble Frontiers Fields
\citep{Montes14, Morishita17, Montes18} or CLASH \citep{Burke15,
DeMaio18} datasets to study ICL at redshifts $z\sim$ 0.3--0.8. Here,
imaging the ICL becomes {\sl even harder} due to cosmological surface
brightness dimming, but through careful background subtraction and
masking of contamination, these studies trace the ICL out to 100 kpc and
beyond, where its rest-frame surface brightness reaches $\mu_V \approx$
26--27 \magsec. Strong color gradients are again seen in the transition
region between the central galaxy and the ICL \citep{Morishita17,
Montes18, DeMaio18}. Stellar population modeling of the ICL colors
indicate populations which are both metal-poor {\sl and} significantly
younger than those in the central galaxy itself \citep{Morishita17,
Montes18}, suggesting that at these redshifts the ICL is actively
growing via the accretion of star forming galaxies. However, there is
still significant uncertainty about when the {\sl bulk} of the ICL
formed; some studies find significant evolution since $z\sim0.5$
\citep{Burke15, Morishita17}, while others infer milder evolution
\citep{Guennou12, Montes18}, and the detection of diffuse light in the
$z=1.2$ cluster MOO J1014+0038 \citep{Ko18} shows that at least some
clusters have a significant ICL component in place quite early.

These studies paint a complex and sometimes contradictory picture where
a variety of mechanisms contribute to ICL formation, including major
mergers, tidal stripping, and galaxy destruction, but the relative
amounts each process contributes, and the timescales on which they
occur, are still only poorly constrained. Several problems contribute to
this muddled picture. One is observational: because of surface
brightness dimming, \mbox{high-$z$} observations often see only the {\sl
brightest} parts of the ICL, and miss a large fraction of the ICL at
lower surface brightnesses. A second problem is definitional: there is
no unambiguous metric for defining ICL, and different studies use
different methods for determining the ICL fraction in clusters. When
these different definitions are applied to simulated clusters \citep[see
the discussion in \citealt{Mihos15}]{Puchwein10, Rudick11}, depending on
the metric chosen, the inferred ICL fraction varies by factors of 2--3
even within an individual cluster. Finally, there is the problem of
progenitor bias: the massive clusters studied at $z\gtrsim0.5$ are
extreme overdensities, and not the progenitor population to local
clusters such as Virgo and Fornax. To properly trace the evolution of
ICL over time and across a range of cluster masses, careful attention
will need to be paid to constructing proper comparison samples at low
and high redshifts.

\subsection{Narrowband Imaging of the Circumgalactic Medium}

Just as deep broadband imaging has revealed the diffuse starlight around
nearby galaxies, deep narrowband imaging has the potential to map the
diffuse circumgalactic medium (CGM) surrounding these galaxies as well.
This warm ionized gas lives at the boundary between galaxies and their
extended environments and traces a variety of physical processes,
including infall from the surrounding environment, the ejection of gas
via AGN or starburst activity, and tidal stripping and shocking of gas.
The CGM has been observed via Ly$\alpha$ emission around bright AGN and
starburst galaxies in ground-based optical imaging and spectroscopy for
high redshift systems, and from space-based UV imaging for objects at
low redshift \citep[see, \eg, ][]{Tumlinson17}. Detecting the CGM around
more quiescent galaxies locally is challenging, however, as it requires
narrowband imaging that is both wide and deep to survey the large
angular scales around nearby galaxies.

\begin{figure}[!h]
\begin{center}
 \includegraphics[width=1.0\textwidth]{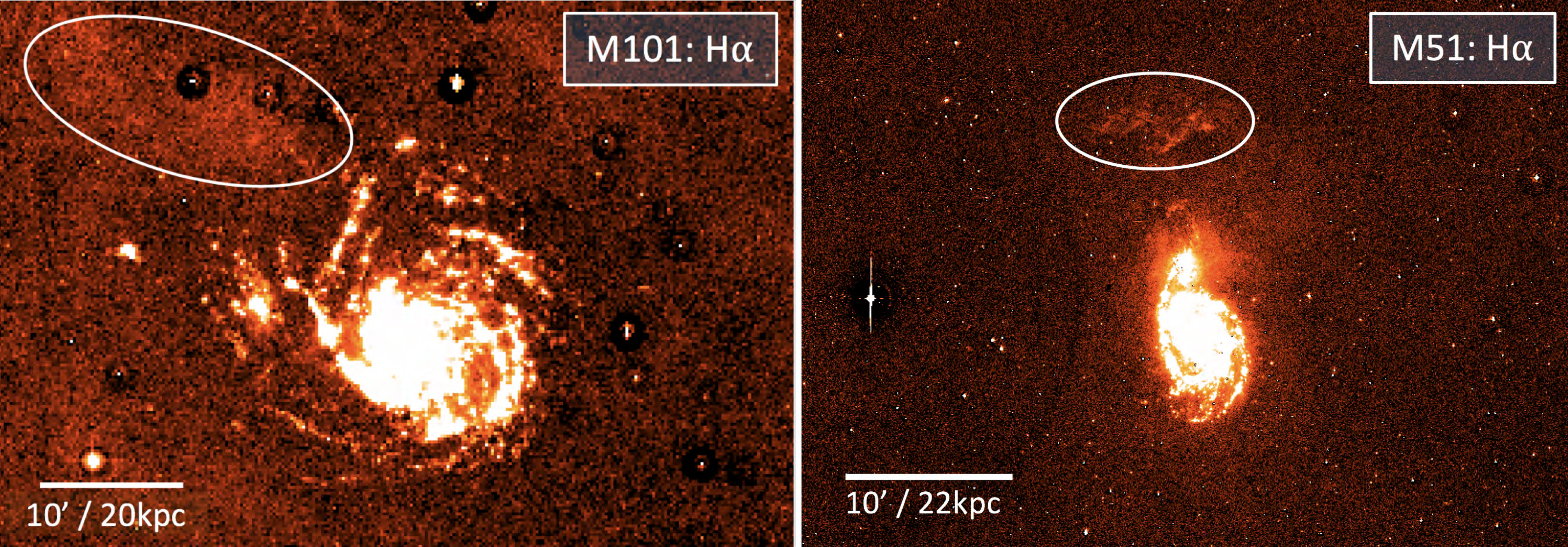} 
 \caption{Deep \ha\ imaging of M101 and M51 from
 \citet[][respectively]{Watkins17,Watkins18}, showing regions of extended
 diffuse \ha\ emission in the outskirts of each system.}
 \label{DeepHa}
\end{center}
\end{figure}

Even in the strongest emission lines, the optical emission from the CGM
is expected to be extremely faint; for example, modeling of \ha\
emission from the CGM at $z=0$ \citep{Lokhorst19} suggests that pushing
down to $\approx 10^{-19}\ {\rm to}\ 10^{-20}$ \haunits\ will be needed
to detect this gas. While this is significantly deeper than current
imaging capabilities, the CGM may have significant substructure
\citep[see \eg\ simulations by][]{Schaye15,vandeVoort19} such that
denser pockets of gas could be detectable at brighter thresholds.
Indeed, this is borne out by recent deep \ha\ imaging of M101 and M51
\citep{Watkins17, Watkins18}, which reach a depth of $5\times10^{-19}$
\haunits\ (see Figure~\ref{DeepHa}). A broad, 40 kpc long plume of
diffuse \ha\ is tentatively detected extending from the northeast edge
of the galaxy's disk, while M51 clearly shows a large ionized cloud just
north of the system. The physical origin of these clouds is unclear,
although their tentative association with tidal features in the galaxies
suggest they may be gas tidally stripped from the disk and shock- or
photo-ionized in the circumgalactic environment.

While observations such as these are intriguing, imaging an order of
magnitude deeper to reach the more diffuse CGM may prove difficult.
Aside from the obvious problem of fewer photons, several technical
challenges arise in doing wide-field narrowband imaging. The fast beams
which deliver the wide field of view and large pixel scale that normally
help with LSB imaging become problematic when coupled with interference
filters, widening the band pass and increasing the sky background. It is
also more difficult to achieve low surface reflectivities on
interference filters, leading to brighter reflections contaminating the
imaging (see \S4.1 below). Working in the red at \ha\ introduces other
problems as well, such as increased sky brightness from OH night sky
lines and contamination due to CCD fringing. However, there are paths
forward. Imaging bluer, at [OIII], alleviates the problems of fringing
and OH lines in the red and results in quieter backgrounds (Mihos \etal,
in prep), although at the lowest metallicities [OIII] emission from the
CGM will be suppressed. Additionally, new optical designs which put the
interference filter at the telescope entrance (Lokhorst, this volume)
could mitigate some of the complications that arise due to fast telescope 
beams. Whether these techniques prove sufficient to reach the truly 
diffuse emission from the extended CGM will be interesting to see.

\section{Challenges at Lower Surface Brightnesses}

Extending the reach of deep LSB imaging to even lower surface brightness
levels will present a number of challenges. The primary sources of
uncertainty are systematic in nature, arising both from instrumental
effects (stray light, flat fielding variations) and from astrophysical
contamination (Galactic cirrus, foreground stars, faint background
sources). While the current LSB imaging techniques mitigate these
effects down to $\approx$ 28--30 \magsec, pushing deeper will require
even more stringent controls on these systematics. Here I highlight a
few of the most concerning issues.

\subsection{Internal Reflections}

Bright objects in the field of view introduce scattered light across the
image, due to the extended wings of the PSF and reflections between the
CCD and various optical elements in the light path. The wings of the PSF
are relatively simple to control, as they are radially symmetric,
centered on the source, and do not vary appreciably as a function of
source position on the image. Deep imaging of bright stars can be used
to measure and model the extended PSF, and correct for its effect on the
images \citep[\eg][]{Slater09, Sandin14, Sandin15, Trujillo16}.
Mitigation of reflections is much more complicated, however.

\begin{figure}[!h]
\begin{center}
 \includegraphics[width=1.0\textwidth]{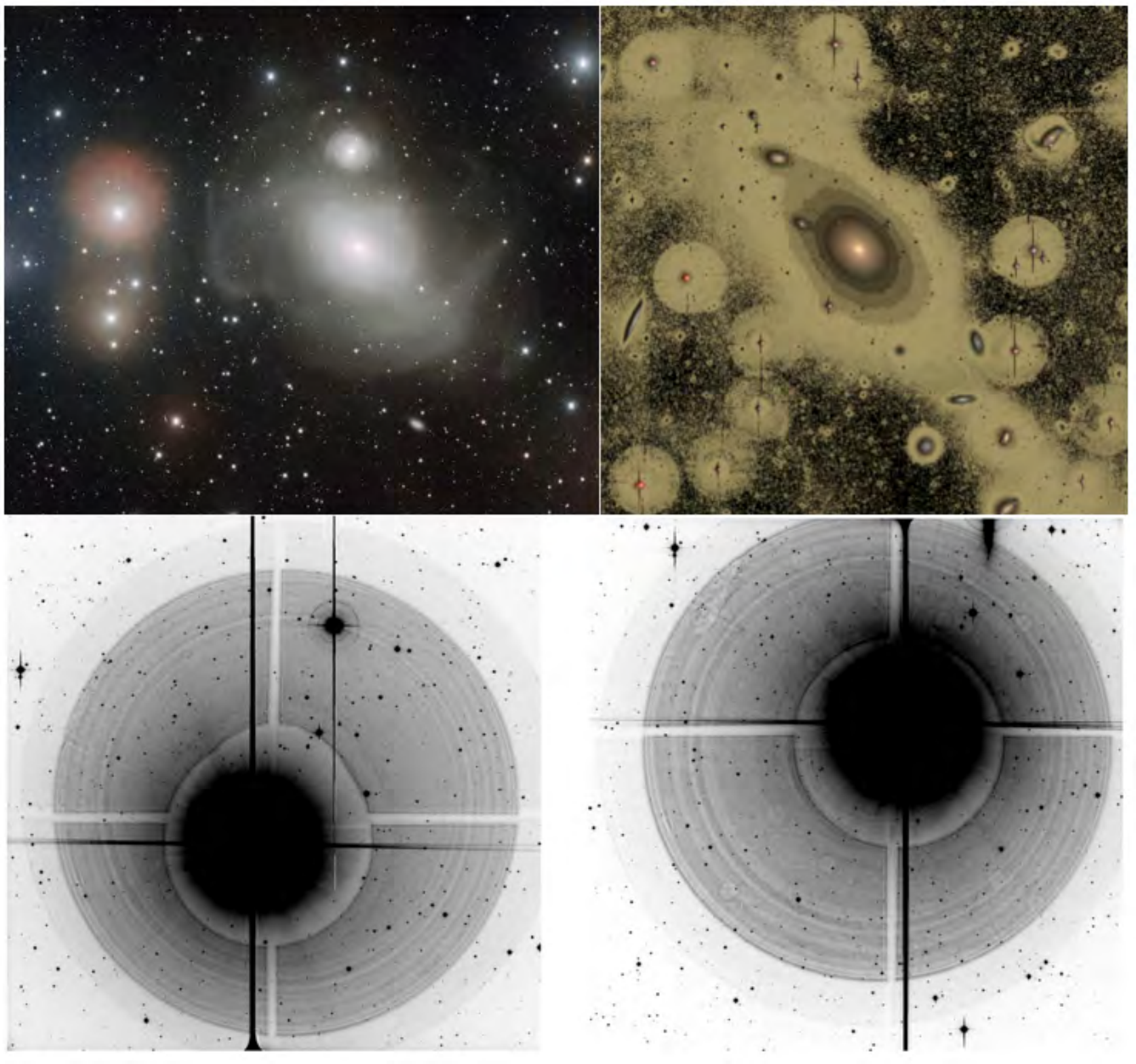} 
 \caption{The impact of internal reflections. Top panels show reflections
 around bright stars in deep imaging from the VST (left) and CFHT (right)
 telescopes. The bottom panels show reflections in a 900s image of
 Arcturus ($V=-0.05$) taken with the Burrell Schmidt telescope
 \citep{Slater09}, showing how the reflections shift when the star is
 moved to different positions in the field. The bright reflections in the
 VST and CFHT data are $\approx 3.5'$ in radius, while the faintest
 reflection visible in the Burrell Schmidt images is 19.5$'$ in radius.}
  \label{reflect}
\end{center}
\end{figure}

Examples of these reflections are shown in Figure~\ref{reflect}. Since
the most reflective surface in any optical camera is the CCD itself,
these reflections arise from light reflecting off the CCD and the back
down from the dewar window, filter, and any other optical element in the
system. These reflections are radially asymmetric, and often have a
complicated nested spatial structure due to bounces between differing
elements. They are also typically off-center with respect to the source,
with a position that changes relative to the source depending on
position in the field. Finally, they have high frequency spatial
structure that is virtually unmodelable, arising from surface variations
on the optical elements, shadowing and scattering off of the secondary
mirror support, and the spatially varying diffraction pattern of the
telescope across the field of view.

These reflection halos around stars can also be quite large ---
extending to tens of arcminutes and beyond --- and imprint complex low
surface brightness patterns over large areas of the image. At current
depths, these reflections can be minimized by the use of aggressive
anti-reflective coatings, reducing the number of reflective elements in
the system, modeling and subtracting reflections, and employing large
scale dither-and-stack techniques to reduce the high frequency residuals
\citep[see \eg][]{Slater09,Karabal17}. However, even in well-designed
optical systems, these reflections become problematic at greater depths.
For example, using the LSST scattered light model \citep{LSSTSci09}, in
LSST imaging, a 7th magnitude star will imprint a 40$'$ diameter halo at
29.5 \magsec, along with reflections stretching across the full
3.5$^\circ$ field of view at 31 \magsec. This is for {\sl one} 7th
magnitude star; the full reflection pattern will be the sum of that
imprinted by {\sl all} stars across the field of view, of which, for
LSST, there will be {\sl many}. Compounding the problem is the fact that
reflections are of course imprinted by all objects in the field,
including the galaxies themselves, which turns the correction problem
into one involving 2D deconvolution rather than simple subtraction. A
full solution for characterizing these reflections and removing this
scattered light from LSST imaging will require a significant investment
of calibration time, computing resources, and work effort if the full
LSB capabilities of the survey are to be realized.

\subsection{Contamination from Galactic Cirrus}

Another significant source of contamination in LSB imaging is the
so-called ``Galactic cirrus'' arising from Milky Way starlight
scattering off dust in the local interstellar medium. In deep optical
imaging, this dust-scattered light creates a patchy and diffuse
foreground screen, which can often mimic tidal structure
\citep{Cortese10}. This dust also radiates thermally in the infrared,
and can be well-traced by deep infared imaging. Cirrus is visible in the
far ultraviolet as well, due to scattering of UV light by the dust
\citep{Witt97,Boissier15}, along with some additional contribution from
molecular hydrogen mixed in with the dust and flourescing in the Lyman
band \citep[\eg][]{Sternberg89, Sujatha10}.

\begin{figure}[!h]
\begin{center}
 \includegraphics[width=1.0\textwidth]{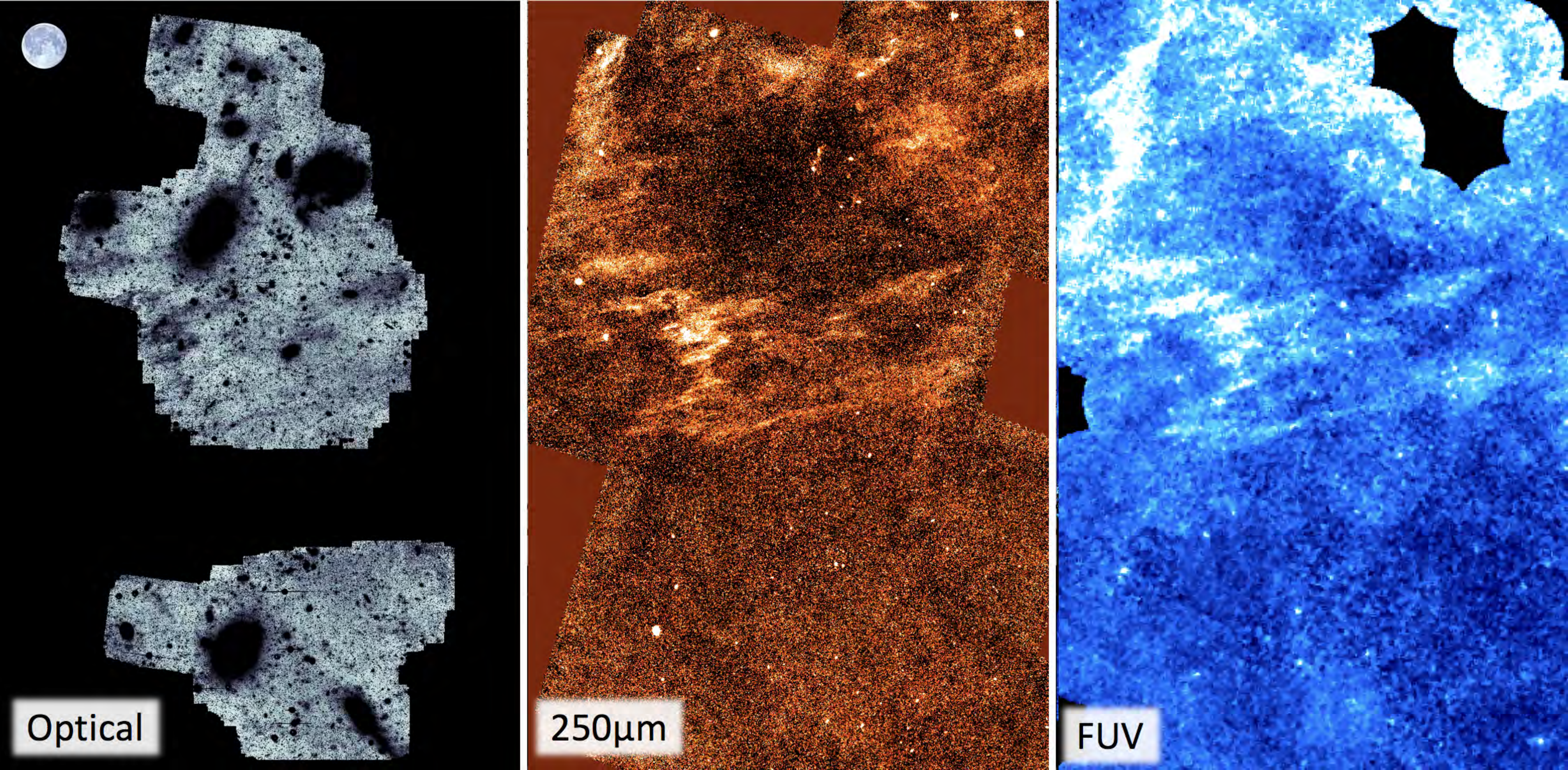} 
 \caption{Deep imaging of the Virgo Cluster in  optical, {\sl Herschel} 250$\mu$m, and
{\sl GALEX} FUV wavelengths \citep[from][respectively]{Mihos17,Davies10,Boissier15}.
The {\sl Herschel} and {\sl GALEX} data reveal emission from Galactic cirrus, tracing
many features also seen in the diffuse optical light. North is up, east is to the left, and
the Moon icon shows a 0.5$^\circ$ scale.}
  \label{VirgoCirrus}
\end{center}
\end{figure}

While Milky Way dust is thickest in the Galactic plane, tenuous cirrus
exists even at at high Galactic latitudes \citep{Planck16} and is easily
visible in deep optical imaging. Figure~\ref{VirgoCirrus} shows imaging
of the Virgo Cluster at optical, {\sl Herschel} 250$\mu$m, and {\sl
GALEX} FUV wavelengths \citep[from][respectively]{Mihos17, Davies10,
Boissier15}. Even at $b=75^\circ$, Virgo has significant contamination
from dust at surface brightnesses of $\mu_V\approx27$ \magsec, visible
in optical imaging as broad and diffuse streamers of light to the north,
south, and southeast of M87 (near the center of the northern part of the
optical image). These features are also seen in both the 250$\mu$m and
FUV imaging, which show M87 and the Virgo core surrounded by a ring of
cirrus. The multiwavelenth data thus acts as a cross-check on the nature
of the diffuse optical light: features that lack FUV/IR emission are
likely to be diffuse starlight around extragalactic sources. M87's
extended stellar halo and the tidal streams to the northwest are all
clear of FUV/IR emission, as are the diffuse light features in the
southern field around M49; all are bona-fide Virgo stellar light
\citep{Mihos17}.

Unfortunately, using the multiwavelength data to make a quantitative
{\sl correction} to optical imaging is much more problematic. There is
rough linear correlation between IR emission and optical surface
brightness in the Galactic cirrus \citep[\eg][]{Witt08}, such that,
given sufficiently deep IR data, one could try a ``scale and subtract''
approach to remove the cirrus \citep{Mihos17}. However, the quality of
this subtraction is quite variable; there is significant scatter in the
IR-optical correlation, since the intensity of the scattered optical
light and the thermal IR emission both depend on a wide range of
parameters, such as the properties of the dust grains, the stellar
populations illuminating the dust, and the relative spatial distribution
of the dust with respect to the illuminating stars
\citep[\eg][]{Bianchi17}. Compounding the problem is the need for deep
and high-spatial resolution IR imaging. While the targeted {\sl
Herschel} 250$\mu$m imaging of Virgo shown in Figure~\ref{VirgoCirrus}
delivers good depth, the 18$''$ beam size is a relatively poor match to
arc-second optical resolution. The situation for other sight-lines
through the Galaxy is worse. One is limited to all-sky far-IR data such
as the reprocessed IRAS 100$\mu$m maps \citep{Miville05} or {\sl Planck}
dust maps \citep{Planck16} which have 4--5$'$ resolution, or the mid-IR
WISE imaging \citep{Meisner14} which has better spatial resolution but
contains a variety of large-scale background residuals that make it less
reliable at the lower surface brightness. Alternatively, one can
attempt to use the FUV maps to trace cirrus, but there can often be
significant differences between the the FUV, IR, and optical maps, as
seen in Figure~\ref{VirgoCirrus}. All these complications make the
multiwavelength data useful as a qualitative signpost for cirrus
contamination, but less helpful for actually removing the contamination.

At lower surface brightnesses the problem is likely to only get worse,
as the sky coverage of Galactic cirrus increases rapidly at more diffuse
levels. Scaling from the \citet{Planck16} dust map, roughly 10\% of the
sky is covered by cirrus with thermal emission similar to that giving
rise to the scattered optical light seen in Figure~\ref{VirgoCirrus}.
The sky coverage increases to 80\% for cirrus with diffuse IR emission a
factor of 20 lower; under a simple linear scaling of the IR and optical
flux, this would correspond to cirrus with an optical surface brightness
of $\mu_V\approx\ $30--31 \magsec. At this depth, with no deep,
high-resolution IR data for guidance, disentangling cirrus contamination
from extragalactic diffuse starlight will be extraordinarily hard. One
possibility is to use the optical data itself: \citet{Roman19} show that
the optical colors of the cirrus in SDSS Stripe 82 imaging are often
distinct from that of extragalactic sources. If this result holds up
more widely, and at even lower surface brightnesses, better
discrimination may be possible. Nonetheless, correction for the 
contamination remains problematic, and cirrus-contaminated fields will
likely remain a significant challenge for LSB science.

\section{Looking Forward}

At the end of our presentations, many of us in the LSB science community
often conclude by saying ``The future of low surface brightness science
is bright!'' While that is most certainly true, I can't also help but
think back the words of \citet{Stebbins34} after tracing M31's surface
brightness so deeply some 85 years ago: ``In fact, were it not for the
interference of the field stars, the detection ... to 27 \magsec\ would
really be easy.'' Over the years, advances in imaging techniques have
allowed astronomers to pierce that threshold, and we are now capable
imaging diffuse light down into the 28--30 \magsec\ regime, with
tantalizing hopes of going even deeper. To echo Stebbins \& Whitford, if
not for the issues of cirrus, scattered light, backgrounds, etc, the
detection to 32 \magsec\ would really be easy. But it won't be easy. We
have not slain these dragons in our modern datasets, we have only driven
them down into the lower depths of the noise. There they linger, lying
in wait for us as we try to push deeper. So to {\sl this} audience I say ``The
future of low surface brightness science is bright, but it's also going
to demand a lot of hard work.'' Now let's get to it.

\acknowledgements 

I would like to thank my long-time collaborator Paul
Harding for his continued technical wizardry with our Burrell Schmidt telescope,
as well as the National Science Foundation, Research Corporation, the Mt
Cuba Astronomical Foundation, and Case Western Reserve University for
their support of our deep imaging work over the years.

\begin{discussion}

\discuss{M. Disney}{How do you envision the future in LSB imaging, both with ground- and space-based instruments?}
\discuss{C. Mihos}{What's that old saying? ``Making predictions is hard, particularly about the future.'' Wide-area deep
surface photometry is {\sl so sensitive} to various systematic effects that going even deeper is going to take a lot
of expensive, pain-staking work in the optical design, scattered light calibration, data reduction techniques, and correction for
backgrounds and foregrounds. It's unclear to me how well we'll be able to do many of these things, or even how {\it committed} 
the broader astronomical community is to investing in this kind of effort. But there are a lot of smart people working
hard on these issues, so I hope to be pleasantly surprised by progress over the coming years...}

\discuss{P.A. Duc}{There is a fair correlation between the WISE 12$\mu$m emission and the cirri which could
help mitigate their effects.}
\discuss{C. Mihos}{At brighter levels, yes, although the WISE imaging has a variety of optical ghosts and other large-scale background 
residuals that make it less reliable at lower surface brightnesses.}

\discuss{D. Valls-Gabaud}{Could an extremely accurate modelling of the full PSF across the FOV, and the ghosts
produced (as done at the CFHT), be helpful in mitigating reflections, or we better design optical configurations which,
by construction, have no reflections other than the ones at the surface of the CCD?}
\discuss{C. Mihos}{An accurate scattered light and PSF model will certainly help, although it can be very expensive in terms
of ongoing calibration as it varies over the field of view and from filter to filter, and also can change over time. And of course
the high frequency spatial  structure of the scattered light will be essentially impossible to model. You are absolutely right 
that optical designs that reduce these effects will be particularly important --- it's always better to eliminate contamination
{\sl before} it gets to the detector, rather than be forced to try and remove it in software after the fact.}

\discuss{S. Driver}{We typically mask the reflections around stars down to quite faint magnitudes. Doesn't this eliminate 
most of the problem?} 
\discuss{C. Mihos}{These reflections can be quite large, and at low surface brightness can cover scales of 0.5$^\circ$ or more
on the imaging. We really don't want to be masking that much of our images as we try to push to lower surface brightness!}

\discuss{J. Murthy}{Some of the FUV emission from the Galaxy appears to arise from H$_2$ fluorescence, not from
scattering from cirri.}
\discuss{C. Mihos}{That's really interesting, and could be part of the reason the structure of the cirrus can sometimes look
significantly different between the FUV and IR imaging.}

\end{discussion}

\end{document}